\documentclass[]{pasj02}
\usepackage[switch,mathlines]{lineno}

\jyear{2026}
\Received{}
\Accepted{}

\graphicspath{{./}{figures/}}

\begin{document}

\title{A pilot submillimeter search for IceCube neutrino counterparts: JCMT follow-up and dusty-galaxy catalog associations}

\author{
 Yuji \textsc{Urata},\altaffilmark{1}\altemailmark\orcid{0000-0002-0786-7307} \email{yjurata@gmail.com}
 and
 Kuiyun \textsc{Huang}\altaffilmark{2}\altemailmark\orcid{0000-0003-2027-372X}\email{kuiyun@gmail.com}
}

\altaffiltext{1}{MITOS Science Co., LTD, New Taipei 235, Taiwan}
\altaffiltext{2}{Center for General Education, Chung Yuan Christian University, Taoyuan 32023, Taiwan}


\maketitle

\begin{abstract}
The submillimeter band offers a complementary route to identifying
electromagnetic counterparts of high-energy neutrinos, tracing both
relativistic transients and persistent dust-obscured galaxies.
We present a pilot search combining JCMT/SCUBA-2 follow-up of IceCube
events with catalog searches for bright high-redshift dusty systems.
Among nine rapidly observed IceCube fields, sources of particular interest
were found in three: JCMT0402$-$0424 (``Shadow Blaster'') in
IC\,210922A, a transient-like source following IC\,220115A, and two known
blazars in IC\,240105A.
ALMA resolved the Shadow Blaster into a strongly lensed dusty star-forming
galaxy at $z=2.988$ with a compact gas-rich starburst core.
Motivated by this result, we cross-matched IceCube localizations with the
Herschel Bright Sources (HerBS) and Planck High-Redshift Source Candidate
(PHZ) catalogs.
Among 51 historical IceCube events with
$A_{\rm bbox,90}\leq\pi~{\rm deg}^{2}$, four fields contain at least one
HerBS source, compared with 0.723 expected from right-ascension scrambling
(fixed-cut $p=4.18\times10^{-3}$).
Scanning the localization-area threshold gives
$p_{\rm global,A}=0.0135$.
Applying the identical IceCube selection to PHZ gives four associated
fields versus 3.544 expected ($p=0.478$).
Because the focused HerBS selection was formulated after examining a
broader exploratory analysis, whose trials-corrected probability is
$p_{\rm global}=0.130$, the focused result is suggestive rather than
definitive.
The four well-localized HerBS fields contain six bright high-redshift
DSFGs with apparent $L_{\rm IR}\sim10^{13}\,L_\odot$.
ALMA imaging and published spectroscopy reveal compact or
multiple-component morphologies, extreme molecular-gas kinematics, and
possible obscured nuclear activity.
These results support a hierarchical strategy: identify luminous
submillimeter candidates through wide-field imaging and catalogs, then
characterize them at high angular resolution to test for compact gas-rich
calorimetric conditions and obscured AGN activity.
\end{abstract}


\section{Introduction}\label{sec:introduction}

The identification of the astrophysical sources of high-energy neutrinos is
one of the central goals of multi-messenger astronomy.
High-energy neutrinos are expected to be produced when accelerated cosmic-ray
protons and nuclei interact with ambient matter or radiation fields,
predominantly through $pp$ or $p\gamma$ interactions.
Because neutrinos propagate essentially unaffected by magnetic fields and
absorption, they provide a direct probe of energetic particle acceleration
even in distant or heavily obscured environments.
However, neutrino observations alone generally provide limited information on
the physical nature, distance, and environment of the source.
Identifying electromagnetic (EM) counterparts to individual neutrino events
is therefore essential for determining the astrophysical sites of cosmic-ray
acceleration and for understanding the origin of the diffuse astrophysical
neutrino flux detected by IceCube
\citep{Aartsen2013Sci,Aartsen2020PRL}.

A major breakthrough was the association of the high-energy neutrino event
IC\,170922A with the flaring $\gamma$-ray blazar TXS~0506+056
\citep{IceCube2018a,IceCube2018b}.
The continuous all-sky monitoring capability of the \textit{Fermi} Large Area
Telescope (LAT) played a key role in identifying the source, which was in an
enhanced $\gamma$-ray state around the neutrino arrival time.
This discovery naturally motivated systematic searches for similar
$\gamma$-ray counterparts to subsequent IceCube alerts.
However, comparable associations have remained rare despite extensive
\textit{Fermi}-LAT and multi-wavelength follow-up campaigns
\citep{Garrappa2024,Li2022}.
Population and stacking analyses also indicate that
$\gamma$-ray--bright blazars can account for only a fraction of the diffuse
astrophysical neutrino flux \citep{Li2022}.

Meanwhile, other source classes have been discussed as possible
high-energy-neutrino emitters, including non-jetted active galactic nuclei
such as NGC~1068, tidal disruption events, and fast blue optical transients
(FBOTs) such as AT2018cow
\citep{ngc1068,Stein2021,2019ApJ...878L..25H,Fang2019}.
AT2018cow is particularly relevant in the present context because its bright
millimeter/submillimeter synchrotron emission provides evidence for strong
non-thermal activity, while theoretical studies have explored FBOTs as
potential sources of high-energy cosmic rays and neutrinos
\citep{2019ApJ...878L..25H,Fang2019}.
These results suggest that a search strategy optimized primarily for
$\gamma$-ray--bright sources may sample only part of the high-energy
neutrino-source population.

This motivates counterpart searches at wavelengths that are sensitive to a
broader range of source classes.
The millimeter and submillimeter bands provide a particularly useful and
complementary window because they can probe two physically distinct
categories of possible neutrino emitters.
First, relativistic transients such as gamma-ray bursts (GRBs), jetted tidal
disruption events, fast blue optical transients, and flaring blazars can be
bright synchrotron sources at millimeter and submillimeter wavelengths.
GRB afterglows in particular have a well-established millimeter/submillimeter
phenomenology
\citep{deUgartePostigo2012,Urata2015}.
A complementary example is the jetted tidal disruption event
Swift~J1644+57, whose rapidly evolving radio-to-submillimeter emission,
including early SMA observations, revealed the formation of a relativistic
outflow
\citep{Zauderer2011}.
As the synchrotron characteristic frequencies evolve, the spectral peak can
pass through these bands on observationally accessible timescales, allowing
the energetics, magnetic field, and relativistic outflow properties to be
constrained.
Millimeter and submillimeter observations can therefore identify and
characterize relativistic sources even when their high-energy emission is
weak, short-lived, or missed by $\gamma$-ray observations.

Second, the same wavelength range is highly sensitive to dust-rich galaxies
that may be difficult to identify in conventional optical, X-ray, or
$\gamma$-ray counterpart searches.
Dusty star-forming galaxies (DSFGs), particularly at
$z\sim1$--4, radiate a large fraction of their bolometric luminosity as
thermal dust emission whose rest-frame far-infrared peak is redshifted toward
the submillimeter bands \citep{Casey2014,Hodge2020}.
Their molecular and atomic emission lines additionally provide redshift and
gas diagnostics.
Compact, gas-rich starbursts are also physically interesting as potential
neutrino sources because efficient confinement of cosmic rays in dense
molecular gas can drive the system toward the calorimetric regime, enhancing
$pp$ interactions and neutrino production
\citep{Thompson2007,Lacki2011,Tamborra2014}.
Thus, submillimeter observations simultaneously provide access to
time-variable relativistic sources and persistent, heavily obscured
galaxies---two source populations that are otherwise usually investigated
using very different observational strategies.

Another practical advantage is the relatively low surface density of bright
submillimeter sources.
At flux densities of several to tens of mJy, the number of unrelated sources
expected within a typical IceCube localization region can be sufficiently
small that a bright detection immediately becomes a useful candidate for
further investigation
\citep{francesco,geach,garratt}.
Motivated by these properties, we initiated a program of wide-field
submillimeter follow-up observations of high-energy neutrino events with the
James Clerk Maxwell Telescope (JCMT) and SCUBA-2.
The large field of view of SCUBA-2 enables a substantial fraction of an
IceCube localization region to be surveyed efficiently, while follow-up
interferometric observations with the Submillimeter Array (SMA) provide
sub-arcsecond localization and multi-frequency measurements required for
counterpart identification and spectral characterization.
This JCMT--SMA strategy was designed without requiring the counterpart to
belong to a specific astrophysical source class.

The effectiveness of this approach was demonstrated by the follow-up of
IC\,210922A.
JCMT/SCUBA-2 observations revealed an exceptionally bright,
persistent submillimeter source, JCMT0402$-$0424 (``Shadow Blaster''), within
the IceCube localization, with an 850-$\mu$m flux density of
$63\pm4$~mJy \citep{GCN30882}.
SMA observations subsequently provided sufficiently accurate localization
for a robust multi-wavelength identification \citep{GCN30891}.
Further observations with ALMA revealed a quadruply lensed DSFG at
$z=2.988$ with a compact, gas-rich starburst core
\citep{Urata2026}.
No comparably plausible $\gamma$-ray, X-ray, or optical transient
counterpart was identified in the IceCube localization.
Together with the low chance-coincidence probability for such an unusually
bright submillimeter source and the physical conditions inferred for its
compact starburst core, JCMT0402$-$0424 was identified as the most plausible
EM counterpart candidate to IC\,210922A
\citep{Urata2026}.
This result demonstrates that neutrino counterpart searches restricted to
the more conventional $\gamma$-ray and optical source populations can miss
candidate sources residing in heavily obscured galaxies.

The discovery of a DSFG in an alert-driven submillimeter search raises the
question of whether similar associations are present among other IceCube
events.
In this paper, we investigate this question using two complementary
approaches.
First, we present the accumulated JCMT/SCUBA-2 follow-up observations of
IceCube events and characterize the bright submillimeter sources identified
within their localization regions.
Second, motivated by the IC\,210922A result, we reverse the search direction
and cross-correlate IceCube events with catalogs of bright, high-redshift
DSFGs.
These event-driven and catalog-driven searches test the same hypothesis from
opposite directions and provide a pilot assessment of the effectiveness of
submillimeter-selected searches for the electromagnetic counterparts of
high-energy neutrinos.

\section{IceCube events and the JCMT sample}
\label{sec:sample}


We compiled high-energy neutrino events followed up with the James Clerk Maxwell Telescope (JCMT) using the SCUBA-2 camera. The observations were carried out through a series of Target-of-Opportunity (ToO) and follow-up programs designed to search for bright submillimeter counterparts within the localization regions of IceCube real-time alerts. 

A practical consideration in the target selection was the relatively large positional uncertainty of individual IceCube events. The localization accuracy varies substantially from event to event, with the 90\% localization regions of track-like alerts commonly extending over angular scales of order $1^\circ$ and, in some cases, considerably larger. We therefore preferentially selected events for which a substantial fraction of the relevant localization region could be mapped with SCUBA-2 to a useful depth within a realistic observing time.

The observing strategy was optimized for relatively bright submillimeter
counterparts.
The rapid observations typically reached 850-$\mu$m rms sensitivities of
a few mJy beam$^{-1}$, with the first-epoch values spanning
$3.2$--$10.5$~mJy~beam$^{-1}$ depending on map size and observing
conditions.
These observations provided sensitivity to sources with flux densities
of order $\sim15$--20~mJy or brighter in the deeper rapid-follow-up
fields.
The achieved sensitivities are sufficient to detect the brighter end of
known millimeter/submillimeter transients, including the jetted tidal
disruption event Swift~J1644+57, AT2018cow, the nearby low-luminosity
GRB\,171205A, and bright classical GRB afterglows such as
GRB\,191221B
\citep{Zauderer2011,2019ApJ...878L..25H,Urata2019,Urata2023}.
Tidal disruption events have been considered promising
high-energy-neutrino source candidates, particularly in light of
candidate associations such as AT2019dsg, while low-luminosity GRBs
have long been discussed as potential high-energy-neutrino emitters
\citep{Stein2021,Murase2006}.
The adopted depth therefore provided useful sensitivity to several
classes of proposed high-energy-neutrino counterparts, while the low
surface density of comparably bright submillimeter sources limited
contamination by unrelated objects.

The JCMT observations can be divided into two complementary modes. The first consists of rapid triggered observations initiated shortly after an IceCube alert. These observations were primarily designed to identify bright submillimeter counterparts and to search for flux evolution indicative of a transient or flaring source. When a promising SCUBA-2 source was identified, interferometric follow-up with the Submillimeter Array (SMA) was used, when possible, to improve the source position from the $\sim14.6''$ SCUBA-2 beam scale to approximately $0.5''$--$2''$ and to obtain additional millimeter/submillimeter spectral information. This coordinated JCMT--SMA strategy enables secure cross-identification with optical, infrared, radio, and X-ray sources. 

The second mode consists of later observations or repeated measurements aimed at identifying persistent submillimeter sources within the neutrino localization regions. These observations became particularly relevant after the discovery of JCMT0402$-$0424 in the IC\,210922A field, where repeated JCMT observations showed no significant variability and subsequent SMA and ALMA observations identified the source as a strongly lensed dusty star-forming galaxy at $z=2.988$ \citep{Urata2026}. The accumulated JCMT sample therefore probes both rapidly varying relativistic counterparts and persistent dust-obscured systems within the same observational framework. 

Table~\ref{tab:too} summarizes the rapid triggered JCMT observations, while Table~\ref{tab:survey} lists the observations used primarily to search for persistent submillimeter sources. In total, the present sample contains 12 IceCube fields observed with JCMT/SCUBA-2. The Gold/Bronze classifications and localization information listed in the tables are taken from the corresponding public IceCube alerts.

\section{JCMT/SCUBA-2 observations and source search}\label{sec:jcmt}

\subsection{Observations and data reduction}\label{ssec:jcmt_reduction}

The IceCube localization regions were observed with SCUBA-2 on the
James Clerk Maxwell Telescope (JCMT), providing simultaneous continuum
imaging at 450 and 850~$\mu$m.
Because the IceCube localization regions typically extend over several
tens of arcminutes, the observations were carried out using the standard
rotating PONG mapping mode.
The PONG size was selected according to the angular extent of each
localization region.
The observations used nominal map diameters of 15, 30, or 60 arcmin,
corresponding to PONG900, PONG1800, and PONG3600, respectively.
The adopted map footprint, first-epoch delay from the IceCube trigger,
and achieved sensitivities are summarized in Table~\ref{tab:too}.

The 850-$\mu$m maps were used as the primary data set for the source
search because the atmospheric transmission, and consequently the
achieved sensitivity, at 450~$\mu$m is substantially more dependent
on weather conditions.
Candidate sources were identified in the matched-filtered
850-$\mu$m maps at a detection significance of $S/N\geq3$,
evaluated using the local rms noise at the source position.
The matched filtering optimizes the detection of beam-sized
point sources while suppressing residual large-scale structure.
The 450-$\mu$m maps were not searched independently for additional
sources.
Instead, they were examined only at the positions of candidates
identified at 850~$\mu$m to determine whether corresponding
450-$\mu$m emission was detected.

The effective SCUBA-2 beam FWHM at 850~$\mu$m is approximately
$14.6''$ \citep{Dempsey2013}.
This is sufficient for identifying bright submillimeter sources within
the IceCube localization regions, but usually insufficient for unique
multi-wavelength identification.
For sufficiently promising candidates, interferometric follow-up with
the Submillimeter Array (SMA) was therefore used, when observationally
feasible, to obtain a more accurate position and additional
millimeter/submillimeter information.

The data were reduced with the \textsc{Starlink/SMURF} pipeline
\citep{Chapin2013}, using the iterative \texttt{makemap} procedure,
including flat-fielding, atmospheric-extinction correction, and removal
of correlated common-mode noise.
Flux calibration was performed using standard JCMT calibrator
observations obtained during the corresponding observing nights,
following the standard SCUBA-2 calibration procedure
\citep{Holland2013}.
The resulting absolute flux calibration accuracy is typically
$\sim5$--10\%.

Because the noise varies across each SCUBA-2 map, the rms values listed
in Tables~\ref{tab:too} and \ref{tab:survey} represent characteristic
map sensitivities over the searched regions, whereas the significance
of an individual candidate was evaluated from the local rms at its
position.

\begin{table*}
\tbl{Summary of the first epoch of triggered JCMT searches.}{%
\begin{tabular}{@{}llrccrrc@{}}
\hline
Event & RA & Dec & Map diameter & Delay &
$450\,\mu{\rm m}$ rms & $850\,\mu{\rm m}$ rms & Candidates \\
      &    &     & (arcmin) & (days) &
(mJy beam$^{-1}$) & (mJy beam$^{-1}$) & \\
\hline
IC200926A (G) & 06:25:50.0 & -04:19:48.0 & 30 & 2.20 &  65.4 &  4.5 & 0 \\
IC200929A (G) & 01:58:07.0 &  03:28:12.0 & 30 & 3.59 &  55.6 &  4.5 & 0 \\
IC210922A (G) & 04:02:55.2 & -04:10:48.0 & 30 & 1.79 &  41.9 &  5.6 & 1 \\
IC220115A (B) & 23:50:15.0 &  26:20:35.0 & 15 & 3.67 &  34.3 &  3.2 & 1 \\
IC220624A (G) & 14:56:28.8 &  41:18:36.0 & 30 & 4.59 &  36.9 &  4.0 & 0 \\
IC221223A (G) & 23:22:09.6 &  34:42:36.0 & 30 & 0.91 &  65.4 &  4.5 & 0 \\
IC230707A (G) & 17:56:07.0 & -01:56:24.0 & 60 & 0.69 & 248.4 & 10.5 & 0 \\
IC230724A (G) & 02:10:04.0 & -01:52:12.0 & 60 & 4.45 & 223.4 &  7.6 & 0 \\
IC240105A (B) & 04:50:00.5 &  11:28:26.6 & 30 & 0.69 &  46.5 &  4.7 & 2 \\
\hline
\end{tabular}}\label{tab:too}
\begin{tabnote}
G and B denote Gold and Bronze IceCube alerts, respectively.
The map diameter corresponds to the nominal PONG mapping footprint:
15, 30, and 60 arcmin correspond to PONG900, PONG1800, and PONG3600,
respectively. Candidate counts refer to sources identified at
$S/N\geq3$ in the 850-$\mu$m maps.
The 450-$\mu$m data were used only to examine emission at the
positions of the 850-$\mu$m candidates.
The quoted rms values are measured from the first-epoch maps.
\end{tabnote}
\end{table*}

\subsection{Triggered-search results}\label{ssec:triggered_results}

The triggered JCMT observations are summarized in
Table~\ref{tab:too}.
Among the nine IceCube fields in the rapid-follow-up sample, bright
submillimeter sources of particular interest were identified in three
fields: IC\,210922A, IC\,220115A, and IC\,240105A.
No comparably significant source was identified in the remaining fields
down to the sensitivities listed in Table~\ref{tab:too}.

The most prominent persistent source was discovered in the field of
IC\,210922A.
JCMT/SCUBA-2 detected JCMT0402$-$0424 (``Shadow Blaster'') with
flux densities of $63\pm4$~mJy at 850~$\mu$m and
$168\pm53$~mJy at 450~$\mu$m \citep{Urata2026}.
Repeated JCMT observations showed no statistically significant variability.
Subsequent SMA and ALMA observations established that the source is a
strongly gravitationally lensed dusty star-forming galaxy (DSFG) at
$z=2.988$, containing a compact, gas-rich starburst core
\citep{Urata2026}.
The detailed multi-wavelength observations and physical interpretation
of JCMT0402$-$0424 have been presented by \citet{Urata2026}; here it is
included as the principal persistent submillimeter counterpart candidate
identified through the JCMT follow-up program.

A qualitatively different candidate was found in the IC\,220115A field.
A point-like 850-$\mu$m source was detected at
RA(J2000) $=23^{\rm h}50^{\rm m}06.559^{\rm s}$ and
Dec(J2000) $=+26^\circ26'31.46''$ in the first SCUBA-2 observation,
obtained 3.67~days after the IceCube trigger.
The peak flux density was 16.9~mJy~beam$^{-1}$, with a local map rms of
3.2~mJy~beam$^{-1}$.

Prompt second-epoch JCMT observations and SMA interferometric follow-up
were planned to test for variability and obtain a more accurate source
position.
However, unfavorable weather prevented sufficiently rapid follow-up.
The field was subsequently re-observed with JCMT on 2022 January 30 and
31, but the source was not recovered in either epoch.
Combining the two late-time observations yields an rms of
3.12~mJy~beam$^{-1}$, corresponding to a $3\sigma$ upper limit of
$S_{850}<9.4$~mJy~beam$^{-1}$.

Subsequent SMA observations were obtained on 2024 August 6 UT
at 225.6 and 347.0~GHz.
No significant counterpart was detected within the positional uncertainty
of the original SCUBA-2 source.
The rms sensitivities were 0.33 and 1.26~mJy~beam$^{-1}$ at
225.6 and 347.0~GHz, respectively, corresponding to $3\sigma$
upper limits of 1.0 and 3.8~mJy~beam$^{-1}$.
The 347.0-GHz observation, which is close in frequency to the
SCUBA-2 850-$\mu$m band, provides a substantially deeper limit
than the flux density measured in the initial SCUBA-2 observation.
The repeated non-detections therefore indicate that the source faded
after the initial SCUBA-2 observation and favor a transient interpretation
over that of a persistent submillimeter source.

\begin{figure}
  \centering
  \includegraphics[width=0.80\columnwidth]{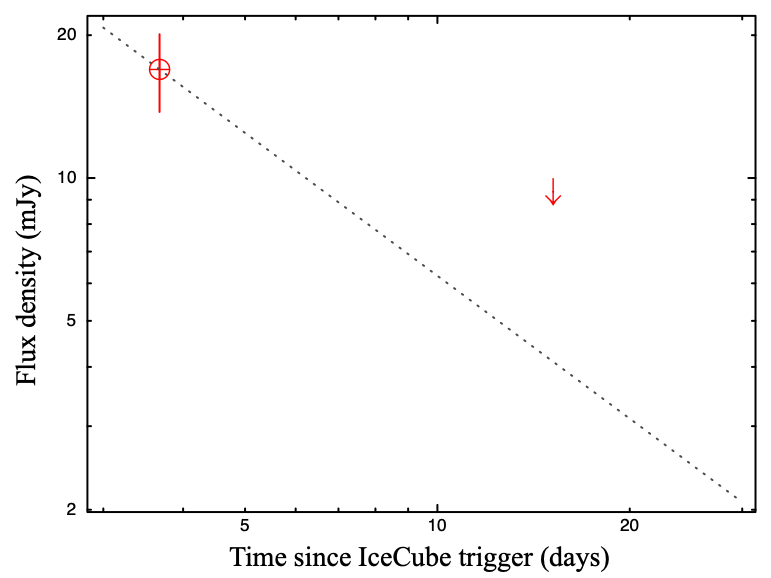}
\caption{
JCMT/SCUBA-2 850-$\mu$m light curve of the transient-like source candidate
in the IC\,220115A localization region.
The source was detected at 3.67 days after the IceCube trigger with a peak
flux density of 16.9 mJy beam$^{-1}$.
The downward arrow shows the $3\sigma$ upper limit from the combined
late-time observations on 2022 January 30 and 31.
The dotted line represents an illustrative power-law decay,
$F_\nu\propto t^{-1}$, normalized to the first-epoch detection, as a
reference for a GRB-afterglow-like temporal evolution.
It is not a fit to the data.
The late-time upper limit is consistent with such a decline, but the sparse
sampling and the lack of prompt interferometric follow-up prevent a
measurement of the decay index or a secure identification of the source.
}
  \label{fig:ic220115a}
\end{figure}

No spatially and temporally coincident GRB trigger was reported by
contemporary high-energy satellite monitors around the time of
IC\,220115A.
This non-detection, however, does not exclude a weak or soft prompt
high-energy transient, such as a low-luminosity GRB, an X-ray-rich
GRB (XRR), or an X-ray flash (XRF), that could have remained below
the sensitivity of the available monitors or outside their effective
coverage.
For reference, Figure~\ref{fig:ic220115a} shows an illustrative
$F_\nu\propto t^{-1}$ decline normalized to the first-epoch SCUBA-2
detection.
The line is intended only as a guide to afterglow-like temporal
evolution and is not a fit to the data.
The subsequent JCMT and SMA upper limits are consistent with such a
decline.

The IC\,240105A field contained two exceptionally bright submillimeter
sources, both identified with previously known blazars, PKS~0446+11 and
J0448+1127 \citep{GCN35499}.
To characterize their short-term behavior, SCUBA-2 observations were
obtained on 2024 January 6, 22, and 28, corresponding to approximately
1, 17, and 23~days after the IceCube trigger.
We measured the source flux densities in the calibrated SCUBA-2 maps
using the point-source fitting procedure for all three epochs.

\begin{figure}
  \centering
  \includegraphics[width=\columnwidth]
  {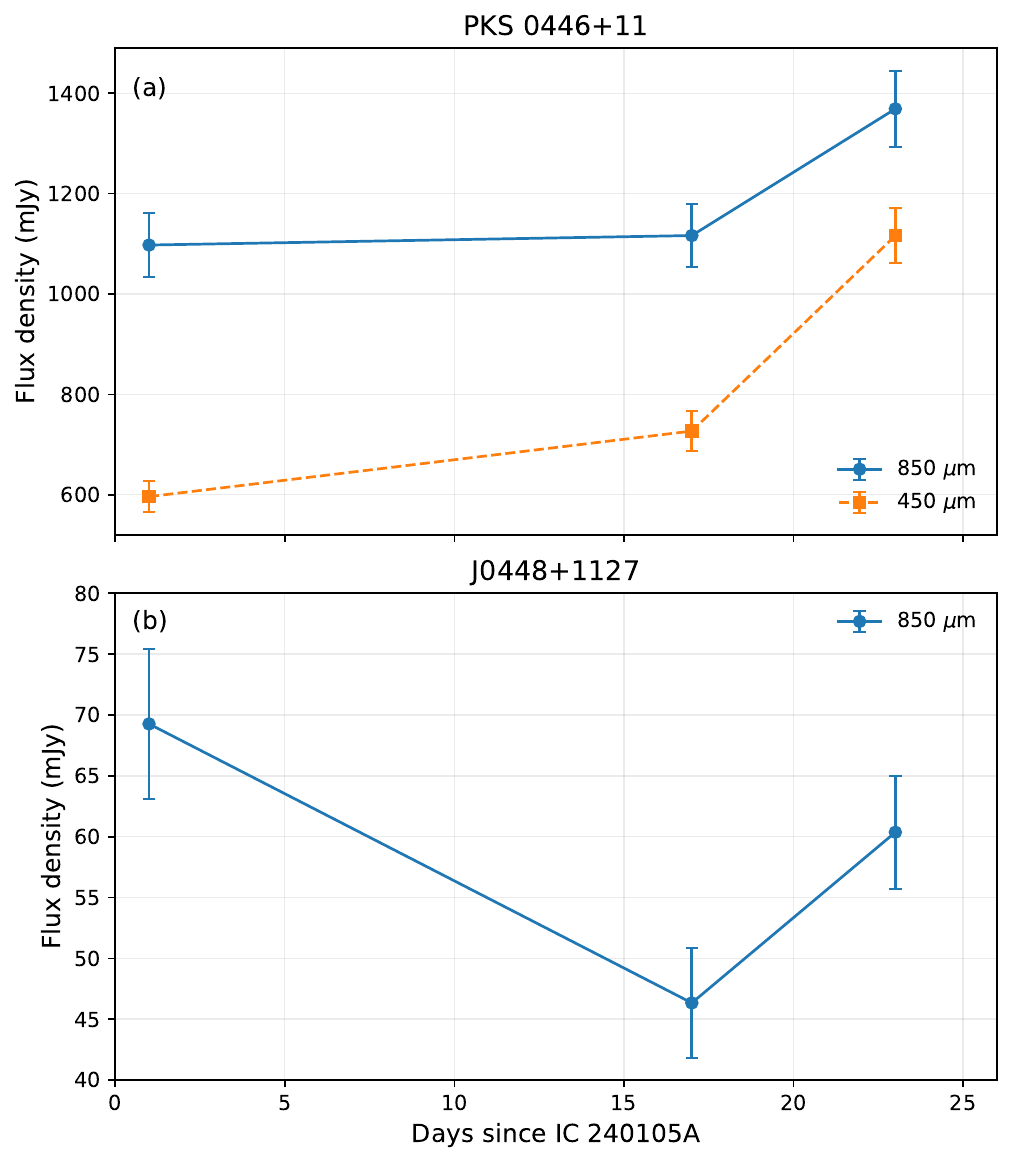}
  \caption{
  JCMT/SCUBA-2 light curves of the two bright blazars in the
  IC\,240105A field.
  (a) PKS~0446+11 at 850 and 450~$\mu$m.
  (b) J0448+1127 at 850~$\mu$m.
  The horizontal axis shows the elapsed time since the IceCube trigger.
  The plotted uncertainties are the formal uncertainties from the
  point-source fits and do not include the absolute SCUBA-2 flux
  calibration uncertainty.
  PKS~0446+11 remained at a similar 850-$\mu$m flux-density level during
  the first two epochs and became brighter by the third epoch; a
  contemporaneous increase is also seen at 450~$\mu$m.
  J0448+1127 shows lower-amplitude variability at 850~$\mu$m.
  }
  \label{fig:ic240105a_lc}
\end{figure}

PKS~0446+11 was detected at high significance in both SCUBA-2 bands
throughout the monitoring sequence.
At 850~$\mu$m, the measured flux densities were
$1.10\pm0.06$, $1.12\pm0.06$, and $1.37\pm0.08$~Jy on
January 6, 22, and 28, respectively, while the corresponding
450-$\mu$m measurements were
$0.60\pm0.03$, $0.73\pm0.04$, and $1.12\pm0.05$~Jy
(Figure~\ref{fig:ic240105a_lc}).
The first two 850-$\mu$m measurements are consistent with a similar
flux-density level, whereas the final epoch is brighter by approximately
20--25\%.
A contemporaneous increase is also present at 450~$\mu$m, where the
change is larger.
The agreement in the direction of the change in both SCUBA-2 bands
suggests genuine submillimeter rebrightening of PKS~0446+11, although
the limited temporal sampling and the larger calibration uncertainty at
450~$\mu$m preclude a precise characterization of the variability.
J0448+1127 was substantially fainter but was detected at 850~$\mu$m in
all three epochs, with measured flux densities of
$69\pm6$, $46\pm5$, and $60\pm5$~mJy, respectively
(Figure~\ref{fig:ic240105a_lc}).
These measurements indicate moderate variability over the three-week
interval, but they do not show a monotonic brightening associated with
the neutrino trigger.
No reliable point-source photometry was obtained for J0448+1127 at
450~$\mu$m.

The JCMT observations therefore do not establish a distinct
submillimeter flaring episode temporally coincident with
IC\,240105A.
These observations do, however, show that the field contained two variable
synchrotron-dominated sources.
PKS~0446+11 remained at a similar 850-$\mu$m flux-density level
during the first two JCMT epochs and showed a possible rebrightening
by the third epoch, while J0448+1127 exhibited more modest variability.
Owing to the sparse temporal sampling and the absence of a pre-trigger
JCMT baseline, these variations cannot be uniquely associated with the
neutrino event.
Nevertheless, the multi-epoch SCUBA-2 observations demonstrate the
value of rapid submillimeter follow-up for distinguishing persistent
or slowly varying jet activity from candidate transient behavior in
neutrino-alert fields.

\subsection{Persistent-source search results}\label{ssec:persistent_results}

In addition to the rapid triggered observations, we carried out later
SCUBA-2 observations of three IceCube fields---IC\,190331A, IC\,201007,
and IC\,220306---with the specific aim of searching for persistent bright
submillimeter sources.
The observations and achieved sensitivities are summarized in
Table~\ref{tab:survey}.
No significant candidate was identified in any of these three fields.

The achieved 850-$\mu$m rms sensitivities were
7.4, 29.9, and 12.8~mJy~beam$^{-1}$ for IC\,190331A,
IC\,201007, and IC\,220306, respectively, corresponding to approximate
$3\sigma$ point-source limits of 22, 90, and 38~mJy~beam$^{-1}$.
The substantial variation in depth reflects both observing conditions and
the large angular extent of the regions that needed to be mapped.

Unlike a rapidly fading transient, a persistent submillimeter counterpart
does not require an observation close to the neutrino arrival time.
Blindly mapping the relatively large IceCube localization regions is
therefore not necessarily the most efficient way to search for this
population, particularly because typical unlensed DSFGs are considerably
fainter than the exceptionally bright JCMT0402$-$0424.
The discovery of JCMT0402$-$0424 consequently motivated a complementary
strategy in which existing catalogs of bright submillimeter-selected
sources are searched directly for positional associations with IceCube
events.
This catalog-based analysis is presented in
Section~\ref{sec:catalog_search}.

\begin{table*}
\tbl{JCMT searches for persistent submillimeter sources.}{%
\begin{tabular}{@{}lccccccc@{}}
\hline
Event & RA & Dec & Map diameter & Epoch & $450\,\mu{\rm m}$ rms & $850\,\mu{\rm m}$ rms & Candidates \\
      &    &     & (arcmin) &       & (mJy beam$^{-1}$) & (mJy beam$^{-1}$)  & \\
\hline
IC190331A & 22:30:43.0 & -20:42:00.0 & 60 & 2023 Sep. & 225 &  7.4 & 0 \\
IC201007  & 17:40:41.0 &  05:20:24.0     & 60 & 2023 Jul. & 934 & 29.9 & 0 \\
IC220306  & 20:59:17.0 &  08:36:36.0     & 60 & 2023 Jul. & 430 & 12.8 & 0 \\
\hline
\end{tabular}}\label{tab:survey}
\begin{tabnote}
Candidate counts refer to sources identified at
$S/N\geq3$ in the 850-$\mu$m maps.
\end{tabnote}
\end{table*}

\section{Catalog searches for bright dusty counterparts}
\label{sec:catalog_search}

The discovery of JCMT0402$-$0424 in the IC\,210922A field showed that
an exceptionally bright persistent submillimeter source can provide a
useful counterpart candidate; the chance probability of finding a source
of comparable 850-$\mu$m brightness in the relevant JCMT search area was
estimated to be $\lesssim1\%$ \citep{Urata2026}.
The rarity of such bright persistent sources motivates the complementary
catalog-based search described here, which extends the search to
pre-selected dusty-galaxy populations over substantially larger survey
areas.

\subsection{Bright submillimeter-selected source samples}
\label{ssec:herbs}

The JCMT follow-up program was designed to identify conspicuous
submillimeter counterparts to individual IceCube events.
Motivated by this strategy, we carried out a complementary
catalog-driven search for bright dusty systems within IceCube
localization regions.
We considered two source populations that probe different parts of
the bright submillimeter regime: the all-sky \textit{Planck}
High-Redshift Source Candidate (PHZ) catalog and the deeper
Herschel Bright Sources (HerBS) catalog.

The \textit{Planck} search is intended primarily as an all-sky search
for exceptionally bright millimetre/submillimetre ``beacons''.
The PHZ catalog contains 2151 high-redshift source candidates selected
from the cleanest 26\% of the sky, with 545-GHz flux densities above
approximately 500~mJy \citep{Planck2016PHZ}.
It therefore probes the extreme bright end of the dusty-source
population.

The Shadow Blaster, JCMT0402$-$0424, discovered in our follow-up of
IC\,210922A, lies in a substantially fainter regime.
It has $S_{850}=63\pm4$~mJy and was subsequently identified as a
strongly lensed DSFG at $z=2.988$ \citep{Urata2026}.
Although the 545-GHz PHZ threshold cannot be compared directly with
an 850-$\mu$m flux density without adopting a spectral energy
distribution, the observed SED of JCMT0402$-$0424 shows that a
Shadow-Blaster-like individual DSFG is not representative of the
extreme flux-density regime systematically selected by PHZ.

Planck is nevertheless demonstrably capable of identifying extreme
strongly magnified DSFGs whose internal properties may be relevant
to the present search.
A well-known example is the population of ``Planck's Dusty GEMS'',
exceptionally bright high-redshift galaxies identified through the
Planck all-sky survey and subsequently characterized with Herschel
and ground-based observations \citep{Canameras2015}.
A particularly well-studied example is
PLCK~G244.8+54.9 (``the Ruby''), a strongly lensed DSFG at
$z\simeq3$.
ALMA observations combined with strong-lensing reconstruction resolve
its molecular gas and dust emission on source-plane scales down to
approximately 60~pc and reveal extremely high-surface-density
star-forming regions, with local star-formation intensities reaching
$\sim2000~M_\odot\,{\rm yr}^{-1}\,{\rm kpc}^{-2}$
\citep{Canameras2017}.
Thus, although Planck selection itself does not establish the presence
of a compact starburst component, it can identify exceptionally bright
lensed systems in which compact and extreme structures are revealed
by subsequent high-resolution observations.

Our primary comparison sample is the Herschel Bright Sources (HerBS)
catalog \citep{Bakx2018}.
HerBS contains 209 bright high-redshift sources selected from the
616.4~deg$^2$ Herschel Astrophysical Terahertz Large Area Survey
(H-ATLAS) \citep{Eales2010HATLAS,Valiante2016HATLAS,Maddox2018HATLAS},
with $S_{500}>80$~mJy and estimated redshifts
$z_{\rm phot}>2$.
At the brightest submillimeter flux densities, magnification bias
makes strongly lensed high-redshift galaxies increasingly prominent
in flux-limited samples, as predicted theoretically and demonstrated
observationally in bright H-ATLAS samples
\citep{Blain1996,Negrello2017}.
Blazar and synchrotron-dominated contaminants were identified using
radio-catalog associations together with the far-infrared-to-submillimeter
spectral energy distributions, including SCUBA-2 measurements, and were
removed from the final sample.
The catalog has also accumulated substantial value-added information from
SCUBA-2 imaging and spectroscopic follow-up
\citep{Bakx2020,Cox2023}.
This makes HerBS a comparatively well-resolved and physically filtered
sample of individual bright high-redshift DSFGs.
The HerBS template-based estimates for the six sources in the four
well-localized IceCube fields are of order
$L_{\rm IR,app}\sim(1$--$3)\times10^{13}\,L_\odot$
\citep{Bakx2018,Bakx2020},
placing them in the apparent hyper-luminous regime.
Such high infrared luminosities are physically relevant because, in
proton-calorimetric starbursts, the neutrino output is expected to scale
approximately with the far-infrared luminosity, which traces the
star-formation-powered cosmic-ray energy budget
\citep{LoebWaxman2006,Thompson2007}.
An empirical correlation between gamma-ray and infrared luminosity in
nearby star-forming galaxies provides complementary evidence that
nonthermal particle output tracks star-formation activity
\citep{Ackermann2012}.
High apparent $L_{\rm IR}$ therefore provides a useful first-stage
selection criterion, although efficient neutrino production additionally
requires compact, gas-rich regions capable of approaching the
calorimetric regime.

HerBS membership alone therefore does not establish the presence of a
compact high-surface-density core, strong gravitational lensing, or an
obscured AGN.
These properties must be determined independently through
high-angular-resolution imaging, spectroscopy, and, for lensed systems,
source-plane reconstruction.
We consequently use HerBS as a candidate-selection sample rather than
as a catalog of pre-identified analogues of the Shadow Blaster.
This motivates the  ALMA and NOEMA characterization presented
in Section~\ref{sec:herbs_characterization}.

IC\,210922A itself lies outside the H-ATLAS footprint from which the
HerBS sample was constructed.
The Shadow Blaster therefore provides the physical motivation and
benchmark for the catalog search, but does not enter the HerBS
cross-match itself.

The Planck PHZ and HerBS samples are consequently complementary.
Planck provides an all-sky search for the rarest and most conspicuous
dusty beacons, including exceptionally bright strongly magnified
systems, whereas HerBS extends the search to fainter individual
high-redshift DSFGs.
Their statistical analyses are therefore performed separately.

\subsection{Cross-match and statistical definitions}
\label{ssec:matching}

For the historical IceCube sample, we selected events with signalness
$\geq0.3$ and excluded events flagged as IceTop-related.
The signalness quantity is the event-level signal proxy reported in the
IceCube alert/catalog framework \citep{Aartsen2017Alerts,Abbasi2023ICECAT}.
The resulting sample contains 226 events spanning
2011 May 14 to 2020 December 22.
IC\,210922A occurred after this period and therefore does not enter the
historical cross-match sample.

For these historical events, publicly available two-dimensional
probability maps are generally unavailable.
We therefore used the asymmetric 90\% localization bounds tabulated
in ICECAT-1 \citep{Abbasi2023ICECAT} to define an approximate
RA--Dec bounding region, whose projected area is denoted by
$A_{\rm bbox,90}$:
\[
A_{\rm bbox,90} =
(\Delta\alpha_{90,+}+\Delta\alpha_{90,-})\cos\delta\,
(\Delta\delta_{90,+}+\Delta\delta_{90,-}),
\]
with all angular quantities expressed in degrees.
All 226 historical events considered here are represented in this way;
none uses an exact MOC localization.
The quantity $A_{\rm bbox,90}$ is therefore an approximate localization
area rather than the area of an integrated two-dimensional 90\% credible
region.

To rank source positions within these asymmetric localizations, we also
define
\[
\rho =
\left[
\left(\frac{\Delta\alpha}
           {\Delta\alpha_{90,\pm}}\right)^2+
\left(\frac{\Delta\delta}
           {\Delta\delta_{90,\pm}}\right)^2
\right]^{1/2},
\]
where the positive or negative 90\% localization bound is selected
independently in each coordinate according to the sign of the source
offset.
The quantity $\rho$ is used only as a normalized positional ranking
statistic and should not be interpreted as a credible level.
For example, $\rho=1$ reaches the normalized boundary along either
coordinate axis, whereas a corner of the corresponding bounding
rectangle lies at $\rho=\sqrt{2}$.

\subsubsection{Focused well-localized sample}
\label{ssec:focused}

The JCMT follow-up strategy preferentially targeted neutrino events with
degree-scale localization regions, for which a substantial fraction of
the relevant uncertainty region could be mapped efficiently with
SCUBA-2 (Section~\ref{sec:sample}).
Motivated by this observational regime, we define a focused
well-localized historical subset using
\[
        A_{\rm bbox,90}\leq\pi~{\rm deg}^{2}.
\]
This threshold is used as a reproducible area-based proxy for
well-localized, degree-scale events rather than as an exact representation
of the JCMT target-selection criterion.
The numerical value $\pi~{\rm deg}^{2}$ is equal in area to a circle of
radius $1^\circ$.
The ICECAT-1 bounding regions are neither circular nor exact integrated
90\% credible regions, and the threshold therefore should not be
interpreted as a circular $1^\circ$ localization radius.
The cut leaves 51 of the 226 historical IceCube events.

For this focused sample we use the simplest field-level statistic:
an IceCube field is counted once if its approximate bounding region
contains at least one catalog source.
Multiple catalog sources in the same IceCube localization therefore do
not increase the number of associations.
We denote this statistic by $N_{\rm field}$.
No additional cut on $\rho$ or on source multiplicity is applied.
The same 51 IceCube events and the same field statistic are used
independently for the HerBS and Planck PHZ catalogs.

Chance expectations were evaluated by right-ascension scrambling,
following the standard approach used in high-energy-neutrino point-source
analyses \citep{Braun2008}.
For each Monte Carlo realization, the declination and localization
geometry of every IceCube event were retained while its right ascension
was randomized.
This preserves the declination-dependent IceCube acceptance and the
localization properties of the observed neutrino sample, while
randomizing its positional relation to the catalog sources.
The catalog positions were kept fixed, thereby retaining the actual
non-uniform sky coverage and clustering of each catalog.
We used $10^{5}$ realizations and applied a plus-one correction to
finite-Monte-Carlo tail probabilities.

As a robustness diagnostic, we repeated the same field-containment test
for the fixed localization-area cuts
\[
 A_{\rm bbox,90}\leq
 1,\ 2,\ \pi,\ 5,\ {\rm and}\ 10~{\rm deg}^{2}.
\]
These values were used to show how the result changes with localization
quality rather than to select the smallest probability.

To quantify the freedom associated with the localization-area threshold,
we additionally scanned $A_{\max}$ over all unique IceCube-event
$A_{\rm bbox,90}$ values between 1 and 10~deg$^{2}$, using the same
field-containment statistic at each breakpoint.
This interval brackets the degree-scale localization regime sampled by
the fixed-area robustness tests.
For each breakpoint, we evaluated the local upper-tail probability.
The identical scan was repeated for every right-ascension-scrambled
realization, and the distribution of the minimum local probability was
used to determine the area-scan-corrected probability
$p_{\rm global,A}$.

This correction accounts only for the freedom to vary
$A_{\max}$ within the adopted 1--10~deg$^{2}$ interval for the
field-containment statistic; it does not include the additional freedom
associated with the choice of test statistic or with the scan interval
itself.
The focused selection was formulated after examination of the broader
exploratory cross-match described in the following subsection.
Its fixed-cut probability is therefore treated as nominal and is not
interpreted as an independently calibrated discovery significance.

\subsubsection{Broader exploratory HerBS analysis}
As a complementary exploratory analysis, we consider the full sample of
226 historical IceCube events.
In this analysis, a HerBS field is selected if it satisfies either a
\textit{central} or an \textit{isolated} criterion.
A central field contains at least one HerBS source satisfying
\[
        \rho_{\rm min}\leq\rho_{\rm cut},
\]
whereas an isolated field satisfies
\[
        A_{\rm bbox,90}\leq A_{\rm cut}
        \quad {\rm and}\quad
        N_{\rm HerBS}=1 .
\]
At the nominal thresholds,
$\rho_{\rm cut}=0.30$ and
$A_{\rm cut}=3.10~{\rm deg}^{2}$.
Only one field is counted irrespective of the number of HerBS sources
inside it.

Because these positional thresholds were exploratory, we calibrated
their look-elsewhere effect with a frozen two-dimensional grid:
\[
        0\leq\rho_{\rm cut}\leq1.00
\]
in steps of 0.05 and
\[
        0\leq A_{\rm cut}\leq6.0~{\rm deg}^{2}
\]
in steps of $0.1~{\rm deg}^{2}$, for a total of 1281 threshold
combinations.
For each scrambled realization, the complete threshold search was
repeated and its minimum local probability was compared with that of the
observed data using the same Monte Carlo ensemble.
This provides the global probability reported below.
Neighboring grid points are strongly correlated; the Monte Carlo
calibration accounts directly for these correlations.

As a secondary robustness test for the exploratory selection, we also
evaluated the signalness-weighted statistic
\[
        T_{\rm sig}=\sum_{i\in{\rm selected}}s_i ,
\]
where $s_i$ is the IceCube signalness of the selected field.

\subsubsection{Contemporary exact-MOC Planck sample}
Finally, we retain an independent contemporary Planck PHZ search using
15 IceCube alerts with public two-dimensional localization maps, spanning
2026 January 15 to August 21.
For these events, Planck positions are tested directly against the
integrated 90\% credible regions of the corresponding multi-order
coverage maps.
Their chance expectation is evaluated by right-ascension shifts of the
MOC regions, preserving their geometry and declination.
This contemporary exact-MOC test is distinct from the historical
well-localized comparison above.

\subsection{Planck PHZ--IceCube associations}
\label{ssec:planck_results}

We first use the Planck PHZ catalog as a comparison population for the
well-localized historical IceCube sample.
Among the 51 events with
$A_{\rm bbox,90}\leq\pi~{\rm deg}^{2}$,
four fields contain at least one PHZ source.
The $10^{5}$ right-ascension-scrambled realizations yield
\[
        N_{\rm obs}^{\rm PHZ,WL}=4,
        \qquad
        \left\langle N_{\rm random}^{\rm PHZ,WL}\right\rangle
        =3.544,
\]
and hence
\[
        p_{\rm nominal}^{\rm PHZ,WL}=0.478.
\]
The Planck PHZ population therefore shows no excess in the same
well-localized IceCube sample used for the HerBS comparison.

The fixed-area robustness checks yield the same qualitative result.
For $A_{\rm bbox,90}\leq2$, 5, and 10~deg$^{2}$, the Planck PHZ
upper-tail probabilities are 0.277, 0.153, and 0.351, respectively;
the $1~{\rm deg}^{2}$ subset contains no observed PHZ-associated field.
Thus, none of the fixed localization-area cuts considered here provides
evidence for a population-level PHZ excess.

We separately examined a contemporary sample of 15 IceCube alerts
with public two-dimensional MOC localizations.
Two PHZ sources lie within their exact 90\% credible regions:
PHz~G109.82+68.15 in the IC\,260712A field and
PHz~G221.53+43.23 in the IC\,260115A field
(Table~\ref{tab:planck_icecube}).

PHz~G109.82+68.15 is particularly noteworthy.
It lies only $\simeq3.95'$ from the best-fit direction of
IC\,260712A, whose exact 90\% localization region has an area of
only $A_{90}=0.682~{\rm deg}^{2}$.
The Planck source position corresponds to an integrated credible
level of 0.130, and the source has a 545-GHz flux density of
$F_{545}=0.856$~Jy.
Thus, although the PHZ population as a whole shows no positional excess,
IC\,260712A--PHz~G109.82+68.15 combines a compact neutrino
localization, a small source--event separation, and a very bright
Planck-selected dusty source, making it a particularly useful target
for detailed millimeter/submillimeter characterization.

The second association, PHz~G221.53+43.23, lies $26.25'$ from the
best-fit direction of IC\,260115A in a substantially larger
90\% localization region with
$A_{90}=6.518~{\rm deg}^{2}$.
Its position corresponds to an integrated credible level of 0.106,
and its 545-GHz flux density is $F_{545}=0.756$~Jy.

For the contemporary exact-MOC sample,
\[
        N_{\rm obs}^{\rm PHZ,MOC}=2,
        \qquad
        \left\langle N_{\rm random}^{\rm PHZ,MOC}\right\rangle
        \simeq2.98,
\]
with
\[
        p_{\rm nominal}^{\rm PHZ,MOC}\simeq0.81.
\]
This independent test is likewise consistent with chance superposition.

The absence of a population-level PHZ excess therefore does not imply
that individual Planck-selected associations are physically
uninformative.
The several-arcminute Planck beam can combine an individual strongly
lensed DSFG, multiple blended dusty galaxies, or a larger-scale
overdensity into a single PHZ entry
\citep{Planck2015Overdensity,Planck2016PHZ}.
Determining which of these configurations applies to a promising
association such as IC\,260712A--PHz~G109.82+68.15 requires
substantially higher angular resolution than Planck provides.
The implications for follow-up and source classification are discussed
in Section~\ref{ssec:future}.

\begin{table*}
\centering
\caption{Planck PHZ sources within the exact 90\% localization regions
of the contemporary IceCube sample.}
\label{tab:planck_icecube}
\footnotesize
\setlength{\tabcolsep}{6pt}
\begin{tabular}{llcccc}
\hline
IceCube event &
Planck source &
Separation &
$A_{90}$ &
Credible level &
$F_{545}$ \\
&
&
(arcmin) &
(deg$^{2}$) &
&
(Jy) \\
\hline
IC260712A &
PHz~G109.82+68.15 &
3.95 &
0.682 &
0.130 &
0.856 \\

IC260115A &
PHz~G221.53+43.23 &
26.25 &
6.518 &
0.106 &
0.756 \\
\hline
\end{tabular}

\vspace{1mm}
\begin{minipage}{0.98\textwidth}
\scriptsize
Note. --- $A_{90}$ is the area of the exact 90\% credible region
derived from the public IceCube multi-order coverage map.
The credible level is the integrated localization probability enclosed
by the contour passing through the Planck source position.
$F_{545}$ is the Planck 545-GHz flux density.
A PHZ entry may represent a single strongly lensed DSFG, a blend of
multiple dusty galaxies, or a larger-scale overdensity within the
Planck beam.
\end{minipage}
\end{table*}

\subsection{HerBS--IceCube associations}
\label{ssec:herbs_results}

Applying the focused well-localized selection
$A_{\rm bbox,90}\leq\pi~{\rm deg}^{2}$ leaves 51 historical IceCube
events.
Four of these fields contain at least one HerBS source, with six HerBS
sources in total (Table~\ref{tab:herbs_icecube}).
Because the statistic is field based, these six sources contribute
four, not six, associations.

The closest positional association occurs in the IC\,160731A field.
HerBS-203/J141827.4$-$001703 lies only
$0.0379^\circ$ ($2.27'$) from the best-fit neutrino direction, with
$\rho=0.074$.
The same IceCube bounding region also contains
HerBS-143/J141810.0$-$003747 at
$\rho=0.497$.
The IC\,180125A and IC\,180410A fields each contain a single HerBS
source: HerBS-139/J134855.6+240745 with $\rho=0.797$, and
HerBS-147/J143403.5+000234 with $\rho=0.729$, respectively.
The fourth field, IC\,141221A, contains two sources:
HerBS-179/J115521.0$-$021329 with $\rho=0.411$ and
HerBS-66/J115820.1$-$013752 with $\rho=0.719$.

The observed field count is therefore
\[
        N_{\rm obs}^{\rm HerBS,WL}=4.
\]
The $10^{5}$ right-ascension-scrambled realizations yield
\[
        \left\langle N_{\rm random}^{\rm HerBS,WL}\right\rangle
        =0.723,
\]
corresponding to the fixed-cut nominal upper-tail probability
\[
        p_{\rm nominal}^{\rm HerBS,WL}
        =4.18\times10^{-3}.
\]
For comparison, the identical 51-event IceCube subset contains four
Planck PHZ-associated fields compared with 3.544 expected randomly
($p=0.478$; Section~\ref{ssec:planck_results}).
The main catalog-association statistics are summarized in
Table~\ref{tab:catalog_statistics}.

The fixed-area robustness checks show that the HerBS excess is not
restricted to a single numerical area cut.
For $A_{\rm bbox,90}\leq2~{\rm deg}^{2}$, two fields are observed
compared with 0.273 expected ($p=0.0280$).
At $5~{\rm deg}^{2}$, four are observed compared with 1.509 expected
($p=0.0598$), and at $10~{\rm deg}^{2}$, seven are observed compared
with 3.008 expected ($p=0.0271$).
The most restrictive $1~{\rm deg}^{2}$ subset contains only nine
IceCube events and no HerBS match.
The trend is therefore not monotonic with localization area, but the
observed HerBS field counts remain above the randomized expectations
over the 2--10~deg$^{2}$ range.
These fixed cuts are nested and strongly correlated and are treated as
descriptive robustness checks rather than as independent trials
(Table~\ref{tab:area_cut_robustness}).

To quantify the freedom associated with the localization-area threshold,
we additionally scanned $A_{\max}$ over all IceCube-event area
breakpoints between 1 and 10~deg$^{2}$, evaluating the same
field-containment statistic at each breakpoint.
The minimum local probability,
$p_{\min}=4.06\times10^{-3}$, is first reached at
$A_{\max}=7.145~{\rm deg}^{2}$.
Applying the identical scan to each right-ascension-scrambled
realization yields an area-scan-corrected probability
$p_{\rm global,A}=0.0135$.
This correction calibrates the freedom to vary the localization-area
threshold within the adopted 1--10~deg$^{2}$ interval for the
field-containment statistic and is not interpreted as a global
significance for the full exploratory analysis.

We also consider the broader exploratory central-or-isolated analysis of
all 226 historical IceCube events.
At the nominal thresholds
$\rho_{\rm cut}=0.30$ and
$A_{\rm cut}=3.10~{\rm deg}^{2}$,
five fields are selected:
IC\,160731A, IC\,161125A, IC\,180125A, IC\,180410A, and
IC\,200921A.
The observed number is five compared with 2.046 expected from randomized
localizations, giving
\[
        p_{\rm nominal}^{\rm HerBS,all}=0.0531.
\]
Across the frozen two-dimensional threshold grid, the minimum local
probability is $p_{\min}=0.0186$.
After calibrating this scan against the scrambled ensemble, the global
probability is
\[
        p_{\rm global}^{\rm HerBS,all}=0.130.
\]
The signalness-weighted statistic yields $p=0.0537$, similar to the
unweighted nominal result.

These three statistical views address complementary aspects of the
same catalog association, but they should not be regarded as independent
tests.
The fixed well-localized selection provides the observationally
motivated field-containment test; the one-dimensional area scan
calibrates the freedom to vary the localization-area threshold within
that statistic; and the broader two-dimensional exploratory analysis
examines a different central-or-isolated statistic over the full
historical sample.
The choice between these statistics is not included in
$p_{\rm global,A}$.

The focused test was formulated after examination of the broader
exploratory analysis, and its small fixed-cut probability is therefore
treated as nominal rather than as an independently calibrated discovery
significance.
The more conservative global probability from the broader exploratory
analysis is $p_{\rm global}=0.130$.

\begin{table*}
\centering
\caption{HerBS sources in the four well-localized IceCube fields.}
\label{tab:herbs_icecube}
\scriptsize
\setlength{\tabcolsep}{3pt}
\begin{tabular}{llllrrrrll}
\hline
IceCube event &
$s$ &
$E_{\rm proxy}$ &
HerBS source &
$A_{\rm bbox,90}$ &
$N_{\rm HerBS}$ &
Separation &
$\rho$ &
Adopted $z$ &
$S_{850}^{\rm SCUBA-2}$ \\
& & (TeV) & & (deg$^{2}$) & & (deg) & & & (mJy) \\
\hline
IC160731A & 0.44 & 98 & HerBS-203 / J141827.4$-$001703 & 1.232 & 2 & 0.0379 & 0.074 & 2.10 phot & $15.0 \pm 4.8$ \\
& & & HerBS-143 / J141810.0$-$003747 & 1.232 & 2 & 0.3320 & 0.497 & 2.2406 spec & $16.4 \pm 3.8$ \\
\hline
IC180125A & 0.36 & 110 & HerBS-139 / J134855.6+240745 & 1.648 & 1 & 0.4401 & 0.797 & 2.54 phot & $15.2 \pm 6.0^{a}$ \\
\hline
IC180410A & 0.60 & 234 & HerBS-147 / J143403.5+000234 & 3.007 & 1 & 0.5173 & 0.729 & 3.1150 spec & $28.2 \pm 4.9$ \\
\hline
IC141221A & 0.35 & 134 & HerBS-179 / J115521.0$-$021329 & 3.028 & 2 & 0.3740 & 0.411 & 3.9423 spec & $32.6 \pm 5.3$ \\
& & & HerBS-66 / J115820.1$-$013752 & 3.028 & 2 & 0.5908 & 0.719 & 2.19 spec & $25.8 \pm 4.2$ \\
\hline
\end{tabular}

\vspace{1mm}
\begin{minipage}{0.98\textwidth}
\scriptsize
Note. --- $s$ is the IceCube signalness and $E_{\rm proxy}$ is the
reported reconstructed energy proxy.
$A_{\rm bbox,90}$ is the approximate projected area of the ICECAT-1
asymmetric 90\% bounding region, and $\rho$ is the normalized
positional-offset statistic defined in Section~\ref{ssec:matching}.
$N_{\rm HerBS}$ is the number of HerBS catalog sources contained in the
same IceCube bounding region.
The field-based statistical test counts each IceCube event only once,
irrespective of $N_{\rm HerBS}$.
The quoted SCUBA-2 flux densities use the revised HerBS measurements of
\citet{Bakx2020}, where applicable, and spectroscopic redshifts are
adopted from the published HerBS/$z$-GAL compilations
\citep{Cox2023}.
$^{a}$HerBS-139 has $S/N\simeq2.5$ in the reprocessed SCUBA-2 map.
This does not affect its inclusion in the present cross-match, which is
defined solely by HerBS catalog membership.
\end{minipage}
\end{table*}

\begin{table*}
\centering
\caption{Statistical summary of the catalog-association searches.}
\label{tab:catalog_statistics}
\footnotesize
\setlength{\tabcolsep}{4pt}
\begin{tabular}{llllcccc}
\hline
Sample & IceCube sample & Localization & Statistic & $N_{\rm obs}$ & $\langle N_{\rm random}\rangle$ & $p_{\rm nominal}$ & $p_{\rm global}$ \\
\hline
HerBS & 51 historical events & $A_{\rm bbox,90}\leq\pi$ deg$^{2}$ & $\geq1$ source per field & 4 & 0.723 & 0.00418 & --- \\
Planck PHZ & same 51 historical events & $A_{\rm bbox,90}\leq\pi$ deg$^{2}$ & $\geq1$ source per field & 4 & 3.544 & 0.478 & --- \\
HerBS & 226 historical events & ICECAT-1 asymmetric bbox & central-or-isolated fields & 5 & 2.046 & 0.0531 & 0.130 \\
Planck PHZ & 15 events, 2026 Jan--Aug & exact 90\% MOC & PHZ-associated fields & 2 & 2.98 & 0.81 & --- \\
\hline
\end{tabular}

\vspace{1mm}
\begin{minipage}{0.98\textwidth}
\scriptsize
Note. --- The first two rows use the identical historical IceCube subset
and the same field-containment statistic, allowing a direct comparison
between HerBS and Planck PHZ.
The cut $A_{\rm bbox,90}\leq\pi~{\rm deg}^{2}$ is equal in area to a
circle of radius $1^\circ$, but the ICECAT-1 bounding regions are not
circular integrated credible regions.
The corresponding $p$ values are fixed-cut nominal probabilities.
The third row summarizes the broader exploratory HerBS analysis; its
$p_{\rm global}$ accounts for the frozen scan over
$\rho_{\rm cut}$ and $A_{\rm cut}$.
The final row is the independent contemporary Planck test using exact
MOC localizations. For the HerBS field-containment statistic, scanning the localization-area
threshold over 1--10~deg$^{2}$ yields
$p_{\rm global,A}=0.0135$; this correction applies only to the freedom
to vary the area threshold within that statistic.
\end{minipage}
\end{table*}

\begin{table*}
\centering
\caption{Fixed localization-area robustness checks for the historical
HerBS and Planck PHZ comparisons.}
\label{tab:area_cut_robustness}
\footnotesize
\setlength{\tabcolsep}{5pt}
\begin{tabular}{rrcccccc}
\hline
$A_{\rm max}$ &
$N_{\rm IC}$ &
$N_{\rm obs}^{\rm HerBS}$ &
$\langle N_{\rm random}^{\rm HerBS}\rangle$ &
$p_{\rm HerBS}$ &
$N_{\rm obs}^{\rm PHZ}$ &
$\langle N_{\rm random}^{\rm PHZ}\rangle$ &
$p_{\rm PHZ}$ \\
(deg$^{2}$) & & & & & & & \\
\hline
1       &   9 & 0 & 0.008 & 1.0000 &  0 &  0.281 & 1.0000 \\
2       &  34 & 2 & 0.273 & 0.0280 &  3 &  1.832 & 0.2765 \\
$\pi$   &  51 & 4 & 0.723 & 0.0042 &  4 &  3.544 & 0.4777 \\
5       &  83 & 4 & 1.509 & 0.0598 & 11 &  7.819 & 0.1528 \\
10      & 128 & 7 & 3.008 & 0.0271 & 18 & 16.197 & 0.3509 \\
\hline
\end{tabular}

\begin{flushleft}
\footnotesize
\textit{Note.} --- Each row uses a fixed upper limit on the approximate
ICECAT-1 bounding-box area and counts IceCube fields containing at least
one catalog source.
The cuts are nested and therefore strongly correlated; they are shown as
descriptive robustness checks and are not treated as independent trials.
For HerBS, an additional scan over all IceCube-event area breakpoints
between 1 and 10~deg$^{2}$ yields a minimum local probability
$p_{\min}=4.06\times10^{-3}$, first reached at
$A_{\max}=7.145~{\rm deg}^{2}$, and an area-scan-corrected probability
$p_{\rm global,A}=0.0135$.
\end{flushleft}
\end{table*}

\section{High-resolution characterization of the HerBS associations}
\label{sec:herbs_characterization}

The focused well-localized cross-match identifies four IceCube fields
containing six HerBS sources.
A positional association alone does not establish which, if any, of these
galaxies produced the corresponding neutrino.
We therefore examined  high-angular-resolution
millimeter/submillimeter observations, published spectroscopy, and
available multiwavelength information to determine whether the associated
fields contain individual compact DSFGs, multiple systems, or unrelated
line-of-sight sources, and to search for possible indications of
obscured AGN activity.

Figure~\ref{fig:herbs_alma} shows  ALMA Band~7 continuum images
of HerBS-203, HerBS-143, HerBS-139, and HerBS-147.
The ALMA continuum images were produced from Band~7 observations
obtained under ALMA project 2024.1.01570.S.
We imaged the calibrated measurement sets with CASA version 6.6.1-17
\citep{CASATeam2022}, using \texttt{tclean} in
multi-frequency-synthesis mode and combining spectral windows 25, 27,
29, and 31 at a representative frequency of approximately 295~GHz.
Briggs weighting with \texttt{robust}=0.5 and a $0.02''$ pixel size
were adopted.
The resulting synthesized beams are typically
$\sim0.14''$--$0.20''\times0.10''$--$0.11''$.
The images are primary-beam corrected, and no gravitational-lens
reconstruction was applied; the descriptions below therefore refer to
the observed image plane.

None of these systems displays the unambiguous four-image or
Einstein-ring morphology characteristic of the Shadow Blaster
\citep{Urata2026}.
Instead, the ALMA images reveal a range of structures from compact
single sources to elongated or multiple-component emission.
HerBS-203 and HerBS-147 are the most suggestive lens or
multiple-component candidates in the present sample, but continuum
morphology alone cannot distinguish gravitational lensing from
mergers, physically associated components, or intrinsic structure.
The comparison therefore illustrates that the HerBS associations are
not simply a population of obvious Shadow-Blaster-like strong lenses.

\subsection{IC\,160731A: HerBS-203 and HerBS-143}
\begin{figure*}[t]
  \centering
  \includegraphics[width=0.93\textwidth]
  {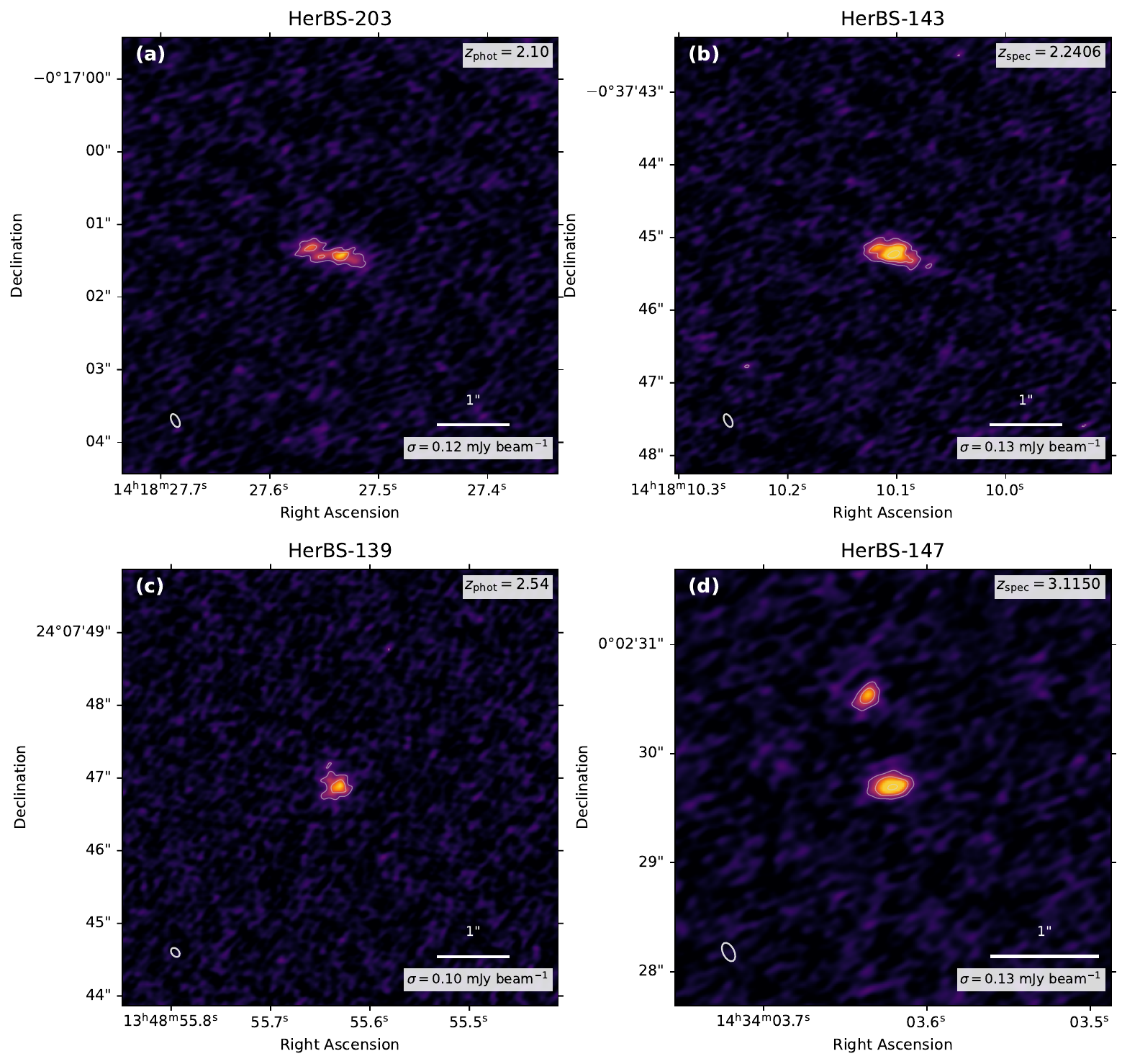}
\caption{
 ALMA Band~7 continuum images of HerBS sources in the focused
well-localized IceCube sample:
(a) HerBS-203 and (b) HerBS-143 in the IC\,160731A field,
(c) HerBS-139 in the IC\,180125A field, and
(d) HerBS-147 in the IC\,180410A field.
The images are centered on the brightest ALMA continuum component.
The field of view is $6''\times6''$ for panels (a)--(c) and
$4''\times4''$ for panel (d).
Contours are drawn at $4$, $8$, and $16\sigma$, with the local rms
indicated in each panel.
The synthesized beam and a $1''$ scale bar are shown in the lower-left
and lower-right regions, respectively.
The adopted photometric or spectroscopic redshift is indicated in each
panel.
None of the four systems shows the unambiguous four-image morphology
of the Shadow Blaster; HerBS-203 and HerBS-147 instead show
morphologies suggestive of lensing or intrinsic multiplicity that require
additional imaging and lens modeling for classification.
}
  \label{fig:herbs_alma}
\end{figure*}

HerBS-203 is the closest HerBS source to an IceCube best-fit direction
in the focused sample, with $\rho=0.074$.
Its ALMA continuum emission is elongated and contains multiple
brightness maxima (Figure~\ref{fig:herbs_alma}(a)), making gravitational
lensing or intrinsic multiplicity plausible interpretations.
However, the continuum morphology alone is insufficient to distinguish
between lensing, interacting components, and intrinsic substructure.

The same IceCube field also contains HerBS-143
(Figure~\ref{fig:herbs_alma}(b)).
Interestingly, HerBS-203 and HerBS-143 have comparable SCUBA-2
850-$\mu$m flux densities,
$15.0\pm4.8$ and $16.4\pm3.8$~mJy, respectively, but their
subarcsecond morphologies are noticeably different:
HerBS-143 is dominated by a more compact continuum component,
whereas HerBS-203 shows a more elongated, multiple-peaked structure.
This comparison illustrates that single-dish submillimeter brightness
alone does not determine whether a bright DSFG is strongly lensed or
intrinsically complex.

Published NOEMA spectroscopy provides
$z_{\rm spec}=2.2406$ for HerBS-143 and detects
CO(3--2), CO(4--3), and [C\,{\sc i}](1--0)
\citep{Cox2023}.
HerBS-203 currently has only a photometric estimate,
$z_{\rm phot}\simeq2.10$.
The two redshifts are therefore compatible at the level allowed by the
photometric uncertainty, raising the possibility that the two bright
DSFGs trace the same broad large-scale environment.
However, their angular separation is of order tens of arcminutes and
HerBS-203 lacks a secure spectroscopic redshift, so we do not claim a
physical association between them.
\subsection{IC\,180125A: HerBS-139}
The IC\,180125A bounding region contains only HerBS-139, making this a
comparatively unambiguous catalog-level association.
HerBS-139 appears as a compact single continuum source in the ALMA image
(Figure~\ref{fig:herbs_alma}(c)).
A WISE source close to the Herschel position has very red mid-infrared
colours, $W1-W2\simeq2.0$ and $W2-W3\simeq3.8$, consistent with the
region occupied by obscured AGNs in commonly used WISE colour diagnostics
\citep{Stern2012,Mateos2012}.
Because the WISE emission is unresolved on the ALMA scale, this remains
suggestive rather than conclusive evidence for an obscured AGN.
A spectroscopic redshift and higher-resolution infrared diagnostics are
needed to establish its physical nature.

\subsection{IC\,180410A: HerBS-147}
The IC\,180410A field likewise contains a single HerBS source.
HerBS-147 is resolved by ALMA into two compact continuum components
(Figure~\ref{fig:herbs_alma}(d)) and has
$z_{\rm spec}=3.1150$.
NOEMA spectroscopy reveals an exceptionally broad molecular-line profile,
with $\Delta V=1789\pm300$~km~s$^{-1}$ in CO(3--2) and CO(5--4),
the largest linewidth in the $z$-GAL sample
\citep{Cox2023}.
The combination of catalog-level isolation, compact multiple components,
and extreme gas kinematics makes HerBS-147 one of the most physically
interesting individual candidates in the focused sample, although the
current data do not distinguish uniquely among a merger, intrinsic
substructure, and gravitational lensing.

\subsection{IC\,141221A: HerBS-179 and HerBS-66}
The IC\,141221A bounding region contains two HerBS sources.
Published spectroscopy provides
$z_{\rm spec}=3.9423$ for HerBS-179 and
$z_{\rm spec}\simeq2.19$ for HerBS-66
\citep{Cox2023}.
Their widely different redshifts show that the two catalog sources do not
form a common physical overdensity.
The multiple occupancy of this IceCube field is therefore a
line-of-sight superposition.
This example also illustrates why the statistical analysis is performed
at the IceCube-field level: the presence of two catalog sources in one
localization is not counted as two independent neutrino associations.

The four fields thus span qualitatively different configurations.
IC\,180125A and IC\,180410A contain one HerBS source each and provide the
cleanest individual catalog-level associations.
IC\,160731A contains two DSFGs whose current redshift information allows,
but does not establish, a common large-scale environment.
IC\,141221A contains two unrelated galaxies projected into the same
neutrino localization.
These distinctions become important when interpreting the population
statistics and when selecting targets for further spectroscopy and
source-plane characterization.

\begin{figure}[tb]
  \centering
  \includegraphics[width=\columnwidth]{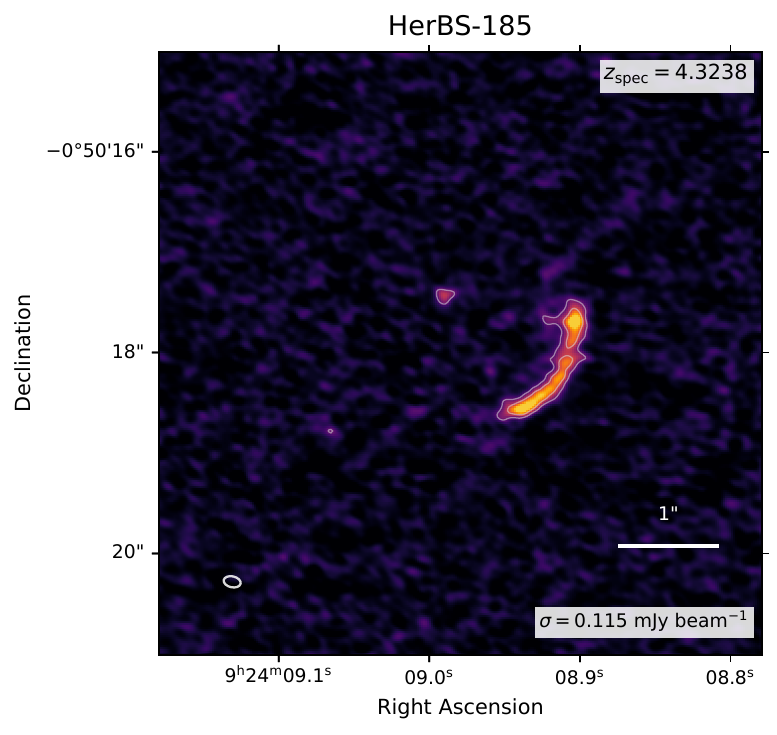}
  \caption{ALMA continuum image of HerBS-185 at
  $z_{\rm spec}=4.3238$ \citep{Cox2023}, in the IC\,161125A field.
  The image shows a prominent arc-like structure in the observed image plane.
  The displayed field is $6''\times6''$.
  Contours start at $4\sigma$ and increase by factors of two,
  where $\sigma=0.115~{\rm mJy\,beam^{-1}}$ is the local rms noise.
  The ellipse at the lower left represents the synthesized beam,
  and the scale bar at the lower right indicates $1''$.}
  \label{fig:herbs185_alma}
\end{figure}

The broader exploratory central-or-isolated selection also contains
the lensed source HerBS-91 \citep{Cox2023} and HerBS-185,
both in the IC\,161125A field.
The ALMA continuum image of HerBS-185 reveals a prominent
arc-like structure suggestive of gravitational lensing
(Figure~\ref{fig:herbs185_alma}).
The IC\,161125A localization has
$A_{\rm bbox,90}=5.377~{\rm deg}^{2}$, exceeding the localization-area
threshold adopted for the focused cross-match.
These objects are therefore discussed separately from the four
well-localized fields considered above.

\section{Discussion}\label{sec:discussion}

\subsection{The role of submillimeter searches in neutrino counterpart identification}
\label{ssec:discussion_strategy}

The JCMT follow-up observations and the catalog-based search demonstrate
two complementary roles of the submillimeter band in identifying possible
electromagnetic counterparts to high-energy neutrinos.
Rapid JCMT observations are sensitive to bright sources without requiring
a prior assumption about source class and can use temporal evolution to
identify transient or flaring candidates.
Persistent DSFGs, by contrast, do not require observations close to the
neutrino arrival time and can be selected efficiently from wide-area
far-infrared and submillimeter surveys.

The well-localized catalog analysis illustrates the connection between
these two approaches.
The JCMT program preferentially targeted events for which a substantial
fraction of the localization region could be mapped to a useful depth
within a realistic observing time.
For the historical IceCube sample, the focused cross-match using the
adopted localization-area threshold yields a low expected number of
chance associations with HerBS sources.
These considerations favor well-localized neutrino events as targets
for both blind submillimeter imaging and catalog-driven searches.

The HerBS and Planck PHZ results further show that catalog selection
cannot be separated from source classification.
HerBS is based on Herschel-resolved bright high-redshift sources and
includes additional filtering and follow-up information, whereas a PHZ
entry is defined at the several-arcminute Planck resolution and may
represent one strongly magnified DSFG, a blend of several dusty galaxies,
or a larger-scale overdensity \citep{Planck2015Overdensity,Planck2016PHZ}.
The absence of a Planck population excess therefore tests the population
of bright Planck-scale dusty structures, not a uniformly classified
population of individual compact DSFGs.

The resulting observational strategy is naturally hierarchical.
Wide-field submillimeter imaging and catalog mining provide the discovery
layer; intermediate- and high-angular-resolution imaging then determine
whether a candidate is a single galaxy, a blend, a lensed system, or an
overdensity; and interferometric spectroscopy and, where appropriate,
gravitational-lens modeling establish the intrinsic source properties.
The IC\,210922A result remains the clearest example of this progression
from single-dish discovery to source-plane physical characterization.

\subsection{Chance associations and statistical interpretation}
\label{ssec:discussion_statistics}

The focused historical comparison provides the clearest indication of a
catalog-level excess.
Four of the 51 IceCube fields with
$A_{\rm bbox,90}\leq\pi~{\rm deg}^{2}$ contain at least one HerBS
source, compared with a randomized expectation of 0.723 fields
($p_{\rm nominal}=4.18\times10^{-3}$).
In contrast, the same 51 IceCube events contain four Planck
PHZ-associated fields compared with 3.544 expected randomly
($p=0.478$).
Thus, under the same IceCube selection and field statistic, an excess is
seen for HerBS but not for Planck PHZ.

The fixed-cut probabilities vary non-monotonically with the adopted
localization-area threshold
(Table~\ref{tab:area_cut_robustness}), indicating some sensitivity
to the choice of $A_{\max}$.
We therefore quantified this freedom with the one-dimensional
area-threshold scan described in Section~\ref{ssec:focused}.
The scan yields a minimum local probability of
$p_{\min}=4.06\times10^{-3}$ at
$A_{\max}=7.145~{\rm deg}^{2}$.
Applying the same scan to the randomized realizations yields
\[
        p_{\rm global,A}=0.0135.
\]
This correction accounts for the freedom to vary the localization-area
threshold within the adopted 1--10~deg$^{2}$ interval for the
field-containment statistic, but not for the choice of test statistic
itself.

The focused excess should nevertheless be regarded as suggestive rather
than as an independent detection.
The focused catalog test was formulated after the broader exploratory
HerBS analysis, for which the full two-dimensional positional-threshold
scan yields
\[
        p_{\rm global}=0.130.
\]
We therefore retain $p_{\rm nominal}=4.18\times10^{-3}$ as the
fixed-cut probability, while using the broader global result as a
conservative measure of the overall statistical evidence.
An independent sample with a prospectively fixed selection will be
required to establish whether the HerBS excess persists.

The limited H-ATLAS footprint also constrains the physical interpretation
of the excess.
Using the actual 616.4~deg$^{2}$ H-ATLAS footprint
\citep{Eales2010HATLAS,Valiante2016HATLAS,Maddox2018HATLAS} and preserving
the declinations of the 51 focused IceCube events yields
\[
        \sum_i Q_i=1.28,
\]
where $Q_i$ is the probability that the randomized best-fit direction
of event $i$ falls within H-ATLAS.
Weighting by the IceCube signalness yields
\[
        \sum_i s_iQ_i=0.72.
\]

In a coupled calculation using the same randomized right ascension
for the H-ATLAS footprint and HerBS chance-coincidence tests, we
adopted a deliberately favorable physical-counterpart model: a
synthetic counterpart is placed at the IceCube best-fit direction
and is assumed to enter the HerBS catalog with unit efficiency
whenever that direction lies within the H-ATLAS footprint.
The field statistic is evaluated as the union of chance associations
and injected counterparts, so that a field satisfying both conditions
is counted only once.

Even when every focused IceCube event is assigned such a HerBS-like
counterpart, the combined chance-plus-physical model yields an expected
number of 1.35 matched fields, with
$P(N_{\rm field}\geq4)=0.039$.
If instead a physical counterpart is assigned with probability equal
to the event signalness, the corresponding expectation is 1.10 fields,
with $P(N_{\rm field}\geq4)=0.020$.
Thus, even under the deliberately favorable direct-counterpart model,
four or more matched fields occur in only a few percent of the
realizations.

These population-level checks therefore do not allow the amplitude of
the excess to be converted directly into a number of genuine
counterparts.
They do, however, make the matched HerBS systems natural high-priority
targets for physical follow-up.
Determining which, and how many, of the associations are genuine
neutrino counterparts requires measurements of the properties most
closely connected to neutrino production, including compact gas
reservoirs, obscured AGN activity, and, where relevant, lensing-corrected
source-plane structure.

The JCMT chance-coincidence estimate for JCMT0402$-$0424
addresses a separate question.
Based on submillimeter source counts, the probability of finding
at least one unrelated source with $S_{850}>50$~mJy within the
IC\,210922A localization was estimated to be $\lesssim1\%$
\citep{Urata2026}.
This probability should not be combined multiplicatively
with the HerBS or Planck catalog probabilities.

\subsection{Physical interpretation of the HerBS candidates}
\label{ssec:physics}

The four well-localized HerBS-associated fields are not physically
homogeneous.
Two, IC\,180125A and IC\,180410A, contain only one HerBS source within
the approximate IceCube bounding region and therefore provide the least
ambiguous individual catalog-level candidates.
The IC\,160731A field contains HerBS-203 and HerBS-143; the spectroscopic
redshift of HerBS-143, $z=2.2406$, is compatible with the current
photometric estimate $z_{\rm phot}\simeq2.10$ for HerBS-203, so a common
large-scale environment remains possible but unconfirmed.
In contrast, the two sources in IC\,141221A have
$z_{\rm spec}\simeq2.19$ and 3.9423 and are clearly unrelated along the
line of sight.
Thus, multiple HerBS sources in one degree-scale neutrino localization
should not by itself be interpreted as a physical overdensity.

There is a second and distinct form of multiplicity on much smaller
angular scales.
ALMA can resolve a single Herschel-selected source into multiple compact
continuum components.
Such structure can arise from gravitationally lensed images, interacting
galaxies, physically associated companions, or intrinsic substructure.
The degree-scale multiplicity of catalog sources and the subarcsecond
multiplicity within an individual HerBS system therefore probe different
physical questions and should not be conflated.

The luminosity-based selection identifies systems with large potential
cosmic-ray power budgets, while the ALMA and lens-modeling analysis tests
whether that power is deposited in sufficiently compact, gas-rich regions
to permit efficient hadronic interactions.
Thus, the physical motivation for considering these DSFGs as possible
high-energy-neutrino environments is not their infrared luminosity alone,
but also the presence of dense regions in which cosmic rays can be
efficiently confined and interact with molecular material.
Such conditions can enhance hadronic $pp$ interactions and drive a
starburst toward the calorimetric regime
\citep{Thompson2007,Lacki2011,Tamborra2014}.
At the population level, however, the accompanying diffuse gamma-ray
emission places important constraints on the fraction of the IceCube flux
that can arise from hadronuclear star-forming populations
\citep{Murase2013,Bechtol2017}.
HerBS-147 is also dynamically notable because its two compact
ALMA components are accompanied by the exceptionally broad
$\Delta V=1789\pm300$~km~s$^{-1}$ molecular-line profile
\citep{Cox2023}.
HerBS-139 instead provides a different possible route, with its compact
continuum morphology and very red WISE colours suggesting a contribution
from a deeply obscured AGN.

Gravitational lensing itself is not an intrinsic signature of neutrino
production.
If the far-infrared- and neutrino-emitting regions are approximately
co-spatial and experience similar magnification, lensing enhances both
observed fluxes by a comparable factor, so the apparent infrared
luminosity remains a useful first-stage proxy for neutrino-source
selection.
This correspondence can be modified by differential magnification if
the emitting regions have different spatial distributions
\citep{Serjeant2012,Hezaveh2012}.

More importantly for the present purpose, strong lensing can
provide the angular magnification needed to reconstruct the
intrinsic source-plane structure of a compact DSFG
\citep{Giulietti2023}.
Combined with molecular-gas measurements, such reconstruction provides
the intrinsic source size and hence the gas surface density, allowing
a direct test of whether the system can approach the cosmic-ray
calorimetric regime.
IC\,210922A demonstrates the importance of this step: lens modeling
revealed a source-plane scale of $\sim0.5$~kpc which, together with the
molecular-gas mass, established a compact, high-surface-density
gas-rich starburst core \citep{Urata2026}.
HerBS-203 and other multiple-component systems identified here therefore
motivate analogous lens tests, molecular-gas measurements, and
source-plane analyses.

The present HerBS associations provide a well-defined set of
targets for extending the investigation initiated with IC\,210922A.
Detailed studies of these systems can establish whether they host
compact, gas-rich starburst cores analogous to that in IC\,210922A
or deeply obscured AGNs.
Determining their intrinsic structures, gas properties, and
dominant power sources through source-by-source analyses is
therefore the essential next step in assessing their physical
plausibility as high-energy-neutrino counterparts.

\subsection{Limitations and future prospects}
\label{ssec:future}

The statistical interpretation is limited by the exploratory origin
of the focused selection and by the restricted H-ATLAS footprint.
Although the one-dimensional area scan yields
$p_{\rm global,A}=0.0135$, this correction accounts only for variation
of the localization-area threshold within the field-containment
statistic, not for the choice among the statistical formulations
considered here.
In addition, even the deliberately favorable direct-counterpart model
gives $P(N_{\rm field}\geq4)=0.039$.
An independent sample with prospectively fixed selection criteria is
therefore required to determine whether the apparent HerBS excess
persists.

The main observational limitation is the incomplete physical
characterization of the candidate systems.
High-angular-resolution ALMA imaging is becoming available for the
HerBS candidates, providing a basis for detailed characterization
of their continuum structure.
For systems affected by strong lensing, lens modeling and
source-plane reconstruction can further constrain their intrinsic
sizes and morphology.
Combining these analyses with molecular-gas spectroscopy and
obscured-AGN diagnostics can establish whether the candidates host
compact gas-rich starbursts or deeply obscured nuclei.
The next step is therefore detailed analysis of the available data
and targeted observations to address the remaining physical questions.

The comparison between HerBS and Planck also emphasizes the importance
of catalog construction and source classification.
The historical Planck PHZ comparison shows no population excess in the
same well-localized IceCube subset, and the contemporary exact-MOC
PHZ test is also consistent with chance.
Individual bright Planck-selected associations nevertheless remain
useful targets for physical follow-up.
PHz~G109.82+68.15 illustrates this point:
the source lies only $3.95'$ from IC\,260712A within the
MOC-defined 90\% localization region of
$A_{90}=0.682~{\rm deg}^{2}$ and has $F_{545}=0.856$~Jy.
The several-arcminute Planck beam, however, does not establish whether
this emission is dominated by a single, potentially strongly lensed
DSFG, a blend of several galaxies, or a larger-scale dusty overdensity
\citep{Planck2015Overdensity,Planck2016PHZ}.
An association confined to a particular physical class could therefore
be diluted in the heterogeneous PHZ population.
HerBS provides more source-level information through its
higher-resolution Herschel selection, contaminant rejection, and
subsequent submillimeter and spectroscopic characterization.

This distinction defines a direct observational next step.
Bright Planck--IceCube positional candidates should first be imaged at
higher angular resolution to determine whether the Planck flux is
dominated by one compact dusty galaxy or by multiple sources.
The individual galaxies identified in these observations can then be
targeted with ALMA, NOEMA, or other interferometers for accurate
localization, molecular and atomic gas spectroscopy, and, where
appropriate, lens identification and source-plane reconstruction.
The aim is to establish whether these fields contain compact gas-rich
cores, dynamically disturbed components, or obscured nuclei that are
physically relevant to neutrino production.

A parallel priority is improved characterization of the four
well-localized HerBS fields.
A spectroscopic redshift for HerBS-203 would test whether the two
IC\,160731A sources could trace a common large-scale environment.
HerBS-139 requires secure spectroscopy and improved obscured-AGN
diagnostics.
HerBS-147 is an especially promising target for detailed dynamical
analysis and tests of gravitational lensing because of its extreme
linewidth and compact multiple components.
The IC\,141221A field provides a useful control example in which
spectroscopy already demonstrates that catalog multiplicity is a
chance line-of-sight superposition.

Future wide-area millimeter and submillimeter facilities can extend
the discovery stage of this strategy.
The Large Submillimeter Telescope (LST), designed for sensitive
wide-field imaging, spectroscopy, and time-domain submillimeter
astronomy \citep{Kohno2023LST}, could identify both dust-obscured
galaxies and variable sources over large sky areas.
The planned all-sky, multi-frequency observations with LiteBIRD
would provide a complementary route to selecting bright
millimeter/submillimeter sources and searching for variability
\citep{LiteBIRD2026}.
Together with the improved localization and increased event
statistics expected from IceCube-Gen2 \citep{Aartsen2021Gen2},
these surveys would enable association tests using larger,
independent samples with prospectively fixed selection criteria.
Such studies would build on both the proposed association of
IC\,210922A with Shadow Blaster \citep{Urata2026} and the
HerBS associations identified in the present analysis.
Combined with detailed characterization of individual candidates,
these tests would assess whether the possible connection between
dusty galaxies and high-energy neutrinos extends to a broader
population and help identify the physical classes most relevant
to neutrino production.

\section{Summary}\label{sec:summary}

We have presented a pilot submillimeter search for electromagnetic
counterparts to high-energy IceCube neutrinos, combining
alert-driven JCMT/SCUBA-2 observations with catalog-based searches for
bright high-redshift dusty galaxies.
The submillimeter band provides a complementary counterpart-search window
because it is sensitive both to time-variable relativistic sources and to
persistent dust-obscured galaxies that may be faint at optical, X-ray, or
$\gamma$-ray wavelengths.

Our JCMT program contains 12 IceCube fields, including nine rapid
triggered observations and three later searches for persistent sources.
Sources of particular interest were found in three triggered fields:
the persistent JCMT0402$-$0424 (``Shadow Blaster'') in IC\,210922A,
a transient-like source following IC\,220115A, and two known blazars in
IC\,240105A.
The Shadow Blaster was subsequently shown by ALMA to be a strongly
gravitationally lensed DSFG at $z=2.988$ containing a compact,
gas-rich starburst core.
Based on submillimeter source counts, the probability of finding
at least one unrelated source with $S_{850}>50$~mJy within
the IC\,210922A localization was estimated to be
$\lesssim1\%$ \citep{Urata2026}.

Motivated by the JCMT preference for well-localized events, we examined
a historical subset of 51 IceCube events satisfying the adopted
localization-area criterion $A_{\rm bbox,90}\leq\pi~{\rm deg}^{2}$.
Four fields contain at least one HerBS source, compared with 0.723
expected from right-ascension-scrambled positions, giving the fixed-cut
nominal probability
$p=4.18\times10^{-3}$.
Applying the identical IceCube subset and field statistic to the Planck
PHZ catalog yields four associated fields, compared with an expectation
of 3.544 from randomized positions ($p=0.478$).
Thus, under the identical IceCube selection and field statistic, the excess
is seen for HerBS but not for Planck PHZ.
A scan over localization-area thresholds yields an
area-scan-corrected probability
$p_{\rm global,A}=0.0135$, which accounts for the freedom to vary
the area threshold within the scanned 1--10~deg$^{2}$ interval.

The focused HerBS result is suggestive but not definitive.
A broader exploratory analysis of all 226 historical IceCube events
selects five central-or-isolated HerBS fields compared with 2.046 expected
($p_{\rm nominal}=0.0531$), and the probability becomes
$p_{\rm global}=0.130$ after calibration of the explored positional
thresholds.
Because the focused well-localized test was formulated after the
broader analysis, its smaller fixed-cut probability is not interpreted
as an independently calibrated discovery significance.
An independent future sample with a prospectively fixed localization
criterion is required.

The four well-localized fields contain six HerBS sources and illustrate
why physical characterization is essential.
IC\,180125A/HerBS-139 and IC\,180410A/HerBS-147 are single-source fields
and provide the least ambiguous individual catalog-level candidates.
IC\,160731A contains HerBS-203 and HerBS-143; their current redshift
information permits, but does not establish, a common large-scale
environment.
IC\,141221A contains HerBS-179 and HerBS-66 at very different
spectroscopic redshifts and is therefore a line-of-sight superposition.
At subarcsecond scales, ALMA further reveals compact and
multiple-component structure, while published spectroscopy identifies
extreme molecular-gas kinematics in HerBS-147 and the multi-wavelength
data suggest possible obscured nuclear activity in HerBS-139.

The Planck null result does not imply that individual bright PHZ systems
lack interest.
Instead, the coarse Planck beam can combine individual lensed galaxies,
blends, and overdensities into a heterogeneous catalog population.
The results therefore support a hierarchical strategy:
wide-field submillimeter discovery and catalog selection,
followed by higher-resolution source decomposition,
interferometric spectroscopy, and, where appropriate, gravitational-lens
reconstruction.
Together with improved IceCube localizations and future wide-area
millimeter/submillimeter surveys, this approach provides a practical path
toward identifying the subset of dust-obscured galaxies capable of
efficient high-energy-neutrino production.



\begin{ack}

We thank the staff of the JCMT, SMA, and ALMA for their support of the
observations.

The JCMT is operated by the East Asian Observatory on behalf of
the National Astronomical Observatory of Japan, Academia Sinica
Institute of Astronomy and Astrophysics, the Korea Astronomy and
Space Science Institute, the National Astronomical Research
Institute of Thailand, and the Center for Astronomical Mega-Science,
with additional funding from the Science and Technology Facilities
Council (UK) and participating universities in the UK and Canada.

The Submillimeter Array is a joint project between the Smithsonian
Astrophysical Observatory and the Academia Sinica Institute of Astronomy
and Astrophysics and is funded by the Smithsonian Institution and the
Academia Sinica.

This study makes use of the following ALMA data:
ADS/JAO.ALMA\#2024.1.01570.S.
ALMA is a partnership of ESO (representing its member states),
NSF (USA) and NINS (Japan), together with NRC (Canada), NSTC
and ASIAA (Taiwan), and KASI (Republic of Korea), in cooperation
with the Republic of Chile. The Joint ALMA Observatory is operated
by ESO, AUI/NRAO and NAOJ.

\end{ack}

\section*{Data availability}

The observational data supporting this study are publicly available
from the respective observatory archives. The JCMT/SCUBA-2 data
are available from the Canadian Astronomy Data Centre (CADC)
under the relevant programme IDs. The SMA data are available
from the SMA Public Archive. The ALMA data are available
from the ALMA Science Archive under project code 2024.1.01570.S.
The IceCube localization data and the catalogs used in the
cross-match analysis are publicly available from the sources cited
in the text.


%

\end{document}